# A three dimensional stochastic Model for Claim Reserving

Magda Schiegl

Cologne University of Applied Sciences
Claudiusstr. 1, D-50678 Köln
magda.schiegl@fh-koeln.de

**Abstract**

Within the Solvency II framework the insurance industry requires a realistic modelling of the risk processes relevant for its business. Every insurance company should be capable of running a holistic risk management process to meet this challenge. For property and casualty (P&C) insurance companies the risk adequate modelling of the claim reserves is a very important topic as this liabilities determine up to 70% percent of the balance sum.

We propose a three dimensional (3D) stochastic model for claim reserving. It meets the necessary number of degrees of freedom to model a realistic claim process that consists of occurrence, reporting and run-off. The model delivers consistently the reserve's distribution function as well as the distributions of all parts of it that are needed for accounting and controlling. The calibration methods for the model are well known from data analysis and they are applicable in an practitioner environment. We evaluate the model numerically by the help of Monte Carlo (MC) simulation.

Classical actuarial reserve models are two dimensional (2D). They lead to an estimation algorithm that is applied on a 2D matrix, the run off triangle. Those methods (for instance the Chain – Ladder or the Bornhuetter – Ferguson method) are widely used in practice nowadays and give rise to several problems: They estimate the reserves' expectation and some of them - under very restriction assumptions - the variance. They provide no information about the tail of the reserve's distribution, what would be most important for risk calculation, for assessing the insurance company's financial stability and economic situation. Additionally, due to the projection of the claim process into a two dimensional space the results are very often distorted and dependent on the kind of projection.
Therefore we extend the classical 2D models to a 3D space because we find inconsistencies generated by inadequate projections into the 2D spaces.

# 1. Introduction

In the framework of Solvency II holistic risk management became very important for insurance business. It is going to implement a new, efficient supervisory basis that enables the risk - orientated and principle based calculation of the economic capital.

In P&C insurance companies the claim reserves are the most important liability position in the balance sheet. They determine up to 70% of the balance sum, of course depending on product mix and capitalisation of the company. Claim reserves are necessary to cover the liabilities arising from insurance contracts written in the presence and the past. Therefore claim reserving is decisive for regulatory and accounting issues as well as for product pricing and reporting.
The claim reserves are calculated for homogeneous portfolios of insurance contracts via actuarial methods which are well known form literature. The basis of the classic reserving methods is built by two dimensional matrices, the run - off – triangles. They are generated via accumulation of claim data. Details are given in section 2 of this paper. An overview on claim reserving and its classical methods can be found in Taylor [12, 13] and in Radtke, Schmidt [7].
The main problem of the classical methods is the fact that a complex stochastic process with many degrees of freedom is transformed to a two dimensional structure – the run-off matrix. This means that "the real world" is projected to a 2D model and it is expected that on this basis one gets a convenient forecast of the future. This is only true under very special assumptions. Additionally most of the classic methods do only estimate the reserves' expectation value. For very few of them, for instance the Chain Ladder method, the second moment of the distribution is know under very restrictive assumptions posed on the underlying stochastic model (see [3]). The point however is that the classical methods give no estimate of the reserve distribution's tail which is of great importance for risk calculation.

In reality the claim process consists of claims occurrence (first), reporting (second) and run off (third). Therefore a 3D model structure is very natural and adequate for this problem. We introduce a 3D stochastic model for claim reserves. The variables of the model are: The number of active claims and the claim payments. The expectation value and the variance of the reserve are given as an analytical function of the model parameters. The reserve distribution can be calculated numerically. To calculate them we perform Monte Carlo (MC) simulations on the basis of this model. There exists already experience on MC simulation techniques for reserve model evaluation, see [8]. The results of the MC simulation are the reserves' probability distributions. The model can be calibrated to real world data (claim portfolio data) via standard methods of data analysis (see [9]). Notice that "reserves" in this context severs as a placeholder for the "total reserve", the "INBR reserve", the "number of IBNR claims" or any other quantity relevant for managing and reporting the reserve process. The distributions are the basis for further risk calculation and management in P&C insurance companies. For instance this makes accessible risk measures as VaR or expected shortfall.

## 2. The connection between the 2D and the 3D models

As was stated in the introduction the classic models project the claim process to a 2D world, operate on cumulated 2D data structures and calculate only the expectation of the reserve. The typical data structures for payments that build the basis for reserve estimation are shown in figure 1. The claim payments are accumulated according to occurrence versus run off years in the one case and reporting versus runoff years in the other case. The formal relation between these two 2D matrices with the tensors of our 3D model is given at the end of this section: $S_{mn}^{(1)}$ are the components of the "occurrence versus run off year" – matrix and $S_{mn}^{(2)}$ are the components of the "reporting versus run off year" – matrix. The matrices' upper left triangles are filled with (known) figures. They represent the past. The figures in the lower right triangle have to be estimated by the application of mathematical, statistical methods. There exists quite a "zoo" of actuarial methods for such calculations. (see for example [1, 2, 3, 6, 7]). Depending on the kind of the used data triangle the result of this calculations is either the total reserve (the sum of reported claims' and IBNR reserves; see left side of figure 1) or two separate results for both parts of the total reserves (see right side of figure 1). In practice one is often faced with the following problem: Appling both methods to the same data set delivers inconsistent results. This is well known and up to now this insufficiency was ascribed to properties of the estimators, particularly the bias (see for instance [10, 11]). To our knowledge up to now nobody asked, if the dimension of the state space underlying the description of the problem is appropriate or not.

The claim process is shown schematically in figure 2. After the occurrence of a claim it is reported to the insurance company. The time between occurrence and reporting, the lag, depends on the line of business and can vary from one day to several years. As an example: For motor car insurance the lag is typically very short, for other lines of third party liability insurance it can be much longer – up to several years. After having been reported the claims are paid – either at once or with several payments. Several payments are typical in the case of injured persons for instance. The time between the reporting of the claim and the final run off can differ from several days up to years or even decades (lifelong annuity payments for injured persons for instance).

Having in mind this three step nature of the claim process (occurrence – reporting – payment) the extension from the classic 2D model to a 3D model is quite natural in the sense that it follows the imperative of reality. We therefore propose as a basis for further analysis to aggregate the claim data to a three dimensional structure: The first dimension is the relative (to the oldest) occurrence year, the second is the number of years between occurrence and reporting and the third dimension is the difference (in years) between claim's reporting and payment.

Claim reserves are modelled stochastically within the above defined 3D framework. It is assumed that the parameters of the "microscopic" claim process are known or can be measured. This is a quite realistic assumption provided the insurance company is ready to collect claim data over a relevant time horizon.

We model a set of claims stochastically, including their projection into the future. We do this by the help of Monte Carlo simulation and get as a result the statistical properties (distribution) of the "macroscopic" quantities (reserves).

We model the following microscopic quantities:

The number of active claims $N_{ijk}$ occurred in year i, reported j years after occurrence and still active k years after reporting. "Still active" means that the claim is not closed and therefore it can be the source of further payments. The second quantity to be modelled is $Z_{ijk}$. This is the total amount paid for the $N_{ijk}$ claims of occurrence year i with reporting delay j years in the kth year after reporting.

In the following we connect the quantities of the 3D model to those of the 2D model which is used by the standard reserving methods. We visualise this circumstances additionally in figure 3.
The up to now reported claims are found in the run off triangle:

$$N_{ij0} \text{ with } i + j \leq I_{max}$$

This triangle (occurrence versus. reporting year) is used in practice to estimate the ultimate number of claims per occurrence year. The boundary $I_{max}$ is given by the claim portfolio's total run-off time (number of years). The estimation of the ultimate claim number consists of the sum of already reported claims and the incurred but not reported (IBNR) claims (see figure 3a). The table of active claims in the presence is given by the set (see figure 3b):

$$\{N_{ijk} | i + j + k = I_{max}\}$$

Below and on this plain lie all known numbers: Number of claims (left side of figure 3) as well as payments for claims(see right hand side of figure 3). The numbers above the plain have to be estimated and give an economic evaluation of the future claim numbers and payments respectively.
The IBNR reserve can be written as:

$$\hat{R}_{IBNR} = \sum_{\{i,j,k | i+j+k > I_{max} \vee i+j < I_{max}\}} \hat{Z}_{ijk}.$$

The number of IBNR claims is given by:

$$\hat{N}_{IBNR} = \sum_{\{i,j | i+j > I_{max}\}} \hat{N}_{ij0}$$

The total reserve evaluates as (sum of reserves for known and unknown claims):

$$\hat{R}_{total} = \hat{R}_{IBNR} + \sum_{\{i,j,k | i+j > I_{max}\}} \hat{Z}_{ijk}.$$

The visualisation of these reserves can be seen in figure 3 on the right hand side: For the IBNR reserve see figure 3d), for the total reserve figure 3c) and d). In both cubes the plain separating the known from the unknown numbers is drawn with sketched lines.
By aggregating the components of the $Z_{ijk}$ tensor (payments) in the following way the run off triangles used by standard 2D reserving methods can be generated:

The run off triangle "payments: occurrence versus run off year", well known in the 2D world, can be generated from the 3D according to:

$$S^{(1)}_{mn} = \sum_{\{j,k \mid j+k=n \vee m+j+k \leq I_{\max}\}} Z_{mjk}$$

This triangle is used in the context of standard reserving methods for estimating the total reserve.

The 2D run off triangle "payment: reporting versus run off year" is generated via:

$$S^{(2)}_{mn} = \sum_{\{i,j \mid i+j=m \vee i+j+n \leq I_{\max}\}} Z_{ijn}$$

This triangle is used in the context of standard reserving methods for business lines with very long reporting times to estimate the total reserve. It should deliver reasonable results in the case of constant portfolio volumes and a stationary claim process.

Insurance companies think currently about setting automatic single claim reserves for some lines of business as for instance motor car insurance. For this purpose they need values for the mean future claim size contingent on the claim history. This automatic procedure does only make sense for claims with small or medium sized expectations. Our model has the advantage to deliver such estimators for the mean claim size depending on the reporting lag j and on the number of years k for which the claim has been active:

$$MCS_{jk} = \frac{\sum_{i,l \geq k} Z_{ijl}}{\sum_i N_{ijk}}.$$

The dependence on k and j is so important because it is expected that the mean claim size grows with both, k and j.

# 3. The 3D Model and its Application

In this section we define the stochastic 3D model. We introduce the necessary distributions determining the single claim process on a "microscopic" level and its parameters. We introduce one example for a realistic claim process and set the parameter values according to this example. This enables us to demonstrate our model in a real world scenario. We implement our model in a Monte Carlo simulation program and calculate the empirical distributions of IBNR claim numbers, IBNR reserve and total reserve.

## 3.1. The 3D Model

As described in section 2 we model the number of active claims $N_{ijk}$ occurred in year i, reported j years after occurrence and still active k years after reporting as well as $Z_{ijk}$, the total amount paid for the $N_{ijk}$ claims of occurrence year i with reporting delay j years in the kth year after reporting.

The ultimate number of claims in occurrence year i is:

$$N_i = \sum_j N_{ij0}$$

It is modelled as Poisson distributed with (given) parameter $\overline{N}_i$.

$$N_i \propto Poisson(\overline{N}_i)$$

The claim numbers obey a multinomial distribution along the years of reporting delays according to the parameters

$$\lambda_j \quad j \in \{1,2,..,I_{max}\} \text{ with } \sum_j \lambda_j = 1.$$

Where the parameter $I_{max}$ is the claim portfolio's total run-off time (number of years). For convenience we have chosen $I_{max} = J_{max}$. This is a suitable assumption if the claim history of the company is known for a long time, minimum as long as the run – off time. Other conventions are possible.

Therefore the claim numbers for k = 0 are distributed according to:

$$N_{ij0} \propto MultNom[N_i, \lambda_1, \lambda_2,..,\lambda_j,..,\lambda_{Imax}]$$

The closing of active claims along the years after reporting is described by the parameters

$$\eta_k \quad k \in \{1,2,..,K_{max}\} \text{ with } \eta_0 = 1 \text{ and } \eta_{k+1} < \eta_k.$$

Where the boundary $K_{max}$ is the maximum number of years a reported claim needs for the run – off to be completed.

The fact that an open claim is not closed in year k can be modelled as a survival process resulting in a binomial distribution with survival probability $\eta_k$:

$$N_{ijk} \propto BiNom[N_{ijk-1}, \eta_k]$$

Therefore the expectation of the active number of claims can be written as:

$$E[N_{ijk}] = \overline{N}_i \cdot \lambda_j \cdot \eta_k.$$

The claim payments are modelled according the collective model of risk theory:

$$Z_{ijk} = \sum_{l=1}^{\nu_{ijk}} X_l^{(ijk)}$$

where the number of single claim payments $\nu_{ijk}$ is binomially distributed

$$\nu_{ijk} \propto BiNom[N_{ijk}, p_{ik}]$$

with parameters $N_{ijk}$, the active number of claims, and $p_{ik}$, the probability that an active claim produces a payment. The $X_{ijk}$ are the payments for single claims. Applying the model to Monte Carlo simulations (section 3.1) we assume them to be Gamma distributed with parameters depending on i, j, and k:

$$X_{ijk} \propto \Gamma[EW_{ijk}, Var_{ijk}]$$

This completes the setting of our model. The choice of the distributions for claim numbers and payments is of course a matter of best fit to the data. We made here the assumptions characterised above, being not too far from real world claim data, in order to apply the model to a concrete simulation example. We deal with some aspects concerning the problem of model selection in section 4.

## 3.2. The Monte Carlos (MC) Simulation and Results

We apply the model described in section 3.1 by the help of MC simulation. We simulate a set of claims in its "microscopic" environment. Every element of the set (i.e. claim) obeys the stochastic process (3D model) of section 3.1. The parameters of the process are known from data analysis of claim databases containing single claim information. The formulas in section 2 give the connection between the microscopic and the macroscopic world, i.e. the reserves for the total claim portfolio. Due to the ansatz of our model we are able to gain knowledge of the macroscopic world's statistical properties, particularly the empirical distributions of the reserves (total and IBNR) and the number of IBNR claims.
We have chosen the model's parameter values in a way that they contain some typical features of real life claim portfolios. They are, of course, no "one to one" values from an existing portfolio.

Now the description of the model parameter values follows: They are displayed in figures 4 and 5. All parameters show typical features as they can be found in third party liability portfolios with person injuries.

The **lag distribution parameters** $\lambda_j$ can be seen in figure 4 on the left side. They are monotonically falling in j, the delay time in years. This is similar in all lines of business, whereas the width of the distribution varies substantially from line to line. In our example some 40% of the claims are reported in the occurrence year and less than 30% one year later.

In figure 4 on the right the **closing down of active claims** is plotted, normalised to 1000 claims at the beginning. From this follow the parameter values for $\eta_k$. Due to the definition it is a monotonically falling function in k, the number of years after reporting. The rate of decrease depends very much of the kind of insurance portfolio. For portfolios with person injuries it is quite typical that even after a very long time – 30 years as in our example or even longer – there are still open claims. This is for instance due to accident – caused, disabled persons that have to be cared for until they die.

For our example we assume that the **probability of payment for an active claim** $p_{ik}$ depends only on k not on i. Active claims, although not fully regulated, do not generate a payment every year. The probability for payment is plotted in figure 5 on the left side. This parameters depends very much on the line of business. In our case there is a rise in probability from a few percent in the early years after reporting up to 80% some 40 years after reporting. In other lines probabilities that decrease with the number of years after reporting can be observed.

For the **expectation value of a single payment** $EW_{ijk}$ we assume homogeneity over occurrence time. Therefore it does not depend on i but on j and k. The expectation value of a single payment is depicted in figure 5 on the right side. The parameter values underlie stylized assumptions as one can see in figure 5. However several typical features are included in this parameter values: The overall size, the fact that there is a maximum in payments some years after the claim was reported and the fact, that this maximum is the higher the longer the time lag between occurrence and reporting was. This can be understood as follows: Claims that are discovered not earlier than a few years after occurrence carry a higher risk than claims having been discovered immediately after occurrence. This is only true for the first few years (in our example 3 years). If the claim is discovered / reported later, the claim size decreases very rapidly. This is of course, one example – in any case the right expectation profile has to be found out via data analysis of the claim portfolio. Further we assumed: $Var_{ijk} = 4EW_{ijk}$. The variance is proportional to the expectation with a constant that is very typical to that kind of claim portfolio. Overdispersed payment distributions are found very often in third party liability insurance.

For our MC simulation we set the initial mean number of claims to $\overline{N}_i = 150$ and assume an annual growth of 3%. This is for example a realistic assumption in the case of an insurance portfolio with 3% growing volume (contracts) and a stationary claim frequency. We considered 15 occurrence years.

For the MC simulation we used the MATHEMATICA package. The results of MC simulation can be seen in figure 6: The distributions for a) the number of claims, b) the IBNR reserve and c) the total reserve. We used 1000 MC samples in this case. In each part of the plot the mean value, depending analytically on the model parameters, is given as a numeric value and is plotted into the graph. From these distributions risk measures as the standard deviation VaR, TailVar or expected shortfall can be calculated. The new regulatory and accounting standards do ask for them.

# 4. Overview and Outlook

This paper proposes a 3D model for claim reserving. It shows the connection to the classical actuarial 2 D models and clarifies that the 3D model can be understood as an extension of the 2D ones. In this way wrong reserve estimations due to inadequate projections on 2D space can be avoided. The 3D model can be calibrated to claim portfolio data via standard methods of data analysis. The expectation and variance of the reserve are an analytical function of the model parameters. The reserve distribution can be calculated by Monte Carlos Simulation techniques. This makes all measures of risk accessible for calculation.
There are many interesting questions that can be investigated as the 3D model implies a new way of analysing claim data:

First of all: What is the performance of 3D models compared with classical actuarial 2D methods. In which situations are 2D methods appropriate? Can quantitative rules be formulated to define regimes in the parameter space where 3D modelling is necessary? Our framework of the 3D models enables us to answer this questions on a quantitative basis. This questions can be investigated via analytical and / or numerical methods. For numerics we are about to set up a MC simulation experiment to compare the 2D and the 3D results for a given "microscopic" claim process according to the model of section 3.1.

Another aim is to put up robust estimation methods for the model parameters from a given claim portfolio. We have already given a first answers to this topic (see [9]): We detected the structure of the claim data set necessary to proceed with the estimation. The estimation has to be consistent in the sense that the estimates of the different parameters do not influence one another. As we have some knowledge of the parameters' properties (see section 3.2) we used a Bayes inference method for parameter estimation. Further practical application would be quite interesting to be done, especially in connection with the points to be investigated as noted above.

After all a further extension of our model to a single claim <u>reserve</u> process would be possible. In an insurance company there are not only claim number and claim payment data available on a single claim level, but also reserve data. The single claim reserves are expert estimations on the final claim size after run off. This estimation is adjusted over the lifetime of the claim to its present state. This could be modelled as a (stochastic) reserve process analogously to the payment process in the present model. The reserve process could also be calibrated on single claim data. The applicability of the extended model in practice has to be analysed carefully due to the following reason: An improved prediction in "macroscopic" portfolio reserve will only be achieved, if the insurance company's real world reserving process can be modelled consistently over the past. All practitioners know that this assumption is not trivial at all! So the success of the model extension depends very much on the peculiarities of the insurance company. Extensions of that kind have been worked out for the 2D methods during the last years (see [5]).

# Literature

1. R. L. Bornhuetter R. E. and Ferguson CAS **59**, 181 (1972).

2. Th. Mack, ASTIN Bulletin **21**, 93 (1991).

3. Th. Mack, ASTIN Bulletin **23**, 213 (1993).

4. Th. Mack, ASTIN Bulletin **30**, 333 (2000).

5. Th. Mack and G. Quarg, Blätter DGVFM Deutsche Gesellschaft für Versicherungs- und Finanzmathematik e.V. – Germany, **XXVI**, 597 (2004).

6. W. Neuhaus, *Scand. Actuar. J.*, 151 (1992).

7. M. Radtke and K.D. Schmidt (eds.), *Handbuch zur Schadenreservierung*, (Verlag Versicherungswirtschaft, Karlsruhe, 2004).

8. M. Schiegl, *On the Safety Loading for Chain Ladder Estimates: A Monte Carlo Simulation Study*, ASTIN Bulletin **32**, 107 (2002).

9. M. Schiegl, *Parameter estimation for a stochastic claim reserving model,* DPG Conference, Berlin (2008).

10. K.D. Schmidt and A. Schnaus, ASTIN Bulletin **26**, 247 (1996).

11. K.D. Schmidt and A. Wünsche DGVM – Blätter **XXIII**, 267 (1998).

12. G.C. Taylor, *Claims Reserving in Non-Life Insurance,* (North-Holland, Amsterdam,1986).

13. G.C. Taylor, *Loss Reserving,* (Kluwer Academic Publishers, Boston, 2000).

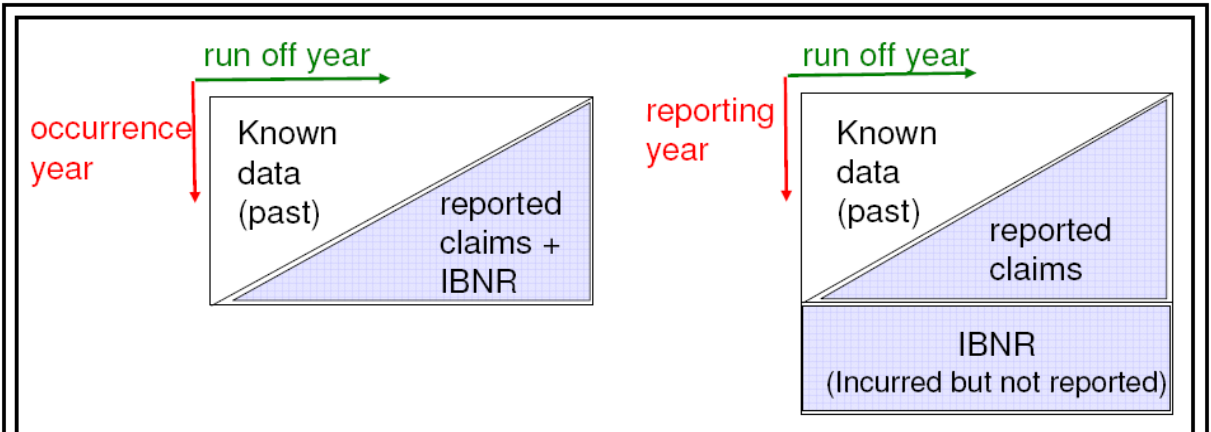

**Fig. 1:**
Typical data structures for reserve estimation.

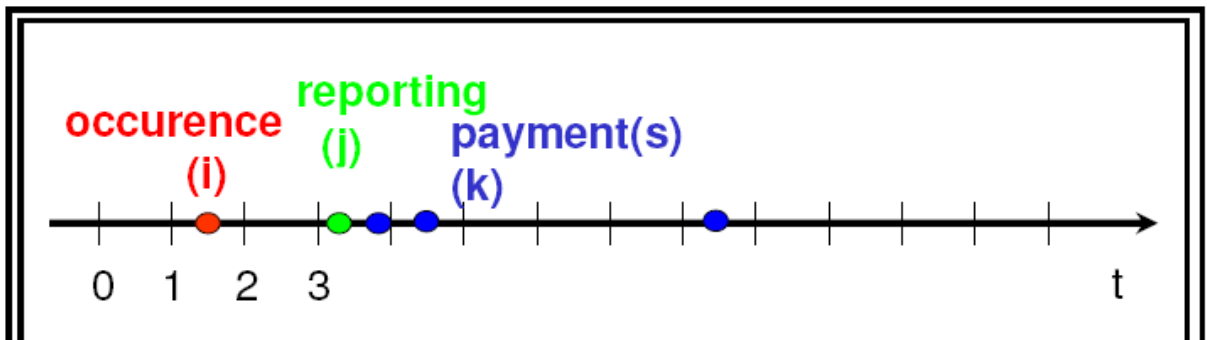

**Fig. 2:**
The claim process: Time diagramm of occurrence, reporting and payment(s)

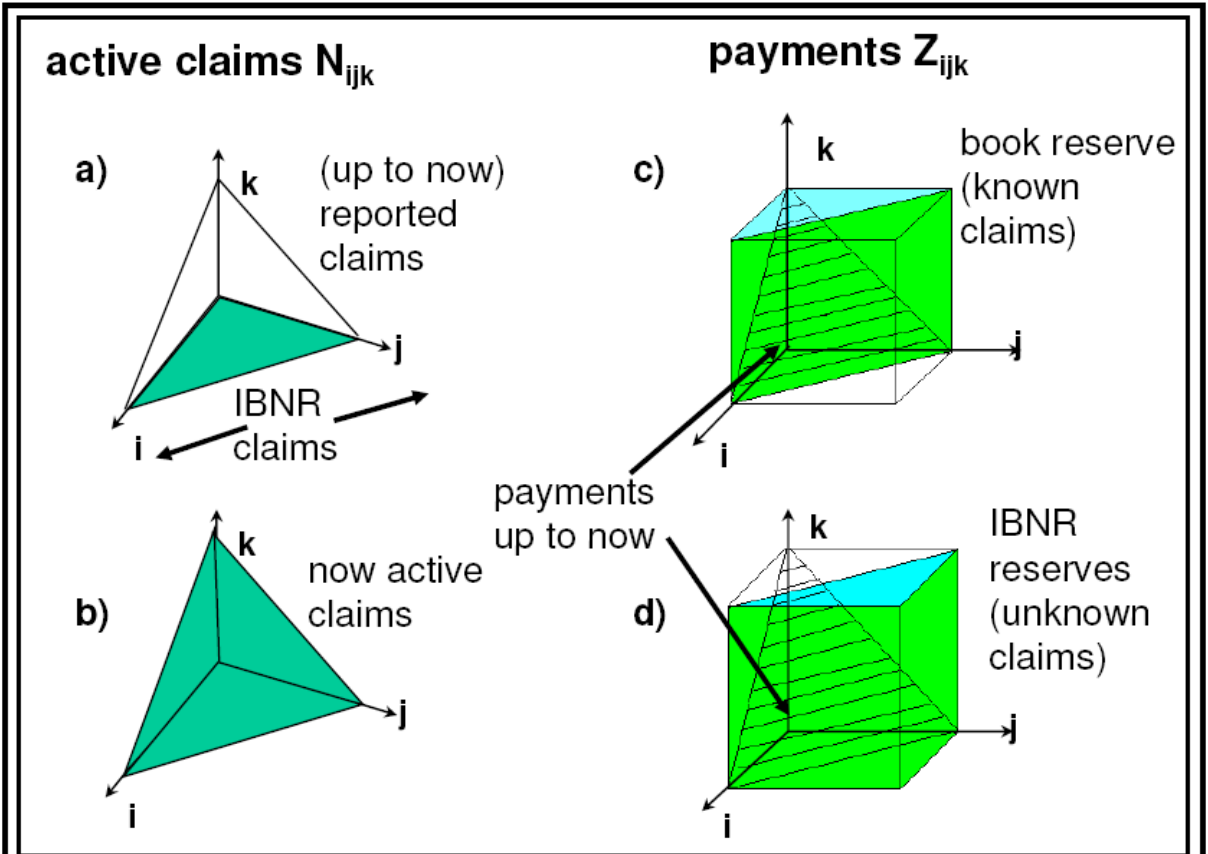

**Fig. 3:**
Visualisation of the 3D model. Left: The active claim numbers; Right: Payments and reserves.

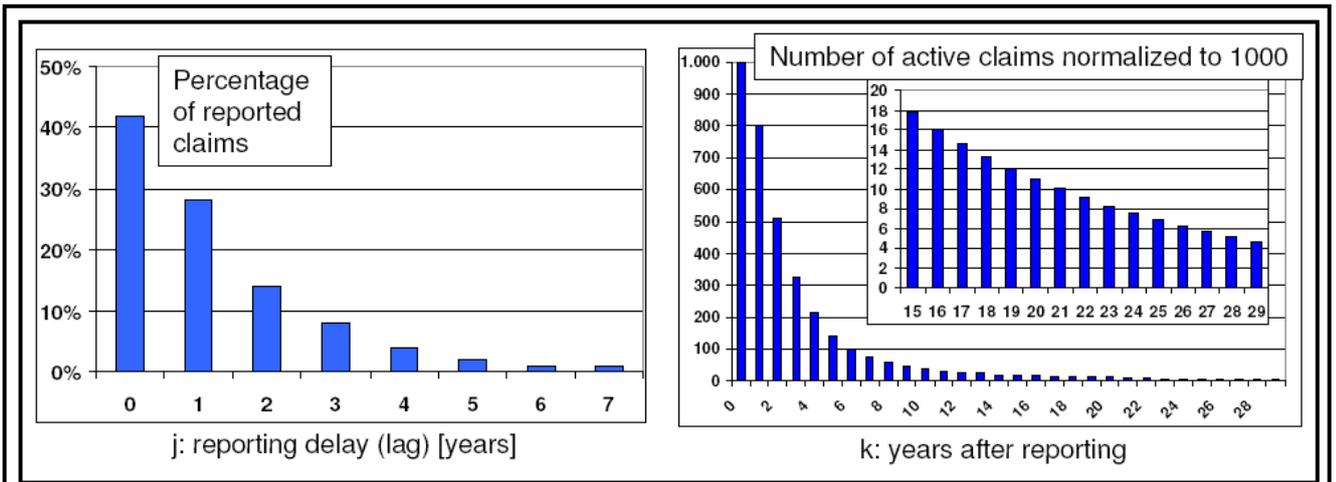

**Fig. 4:**
Model parameters: Lag distribution (left) and Closing of active claims (right) for third party liability insurance / person injuries

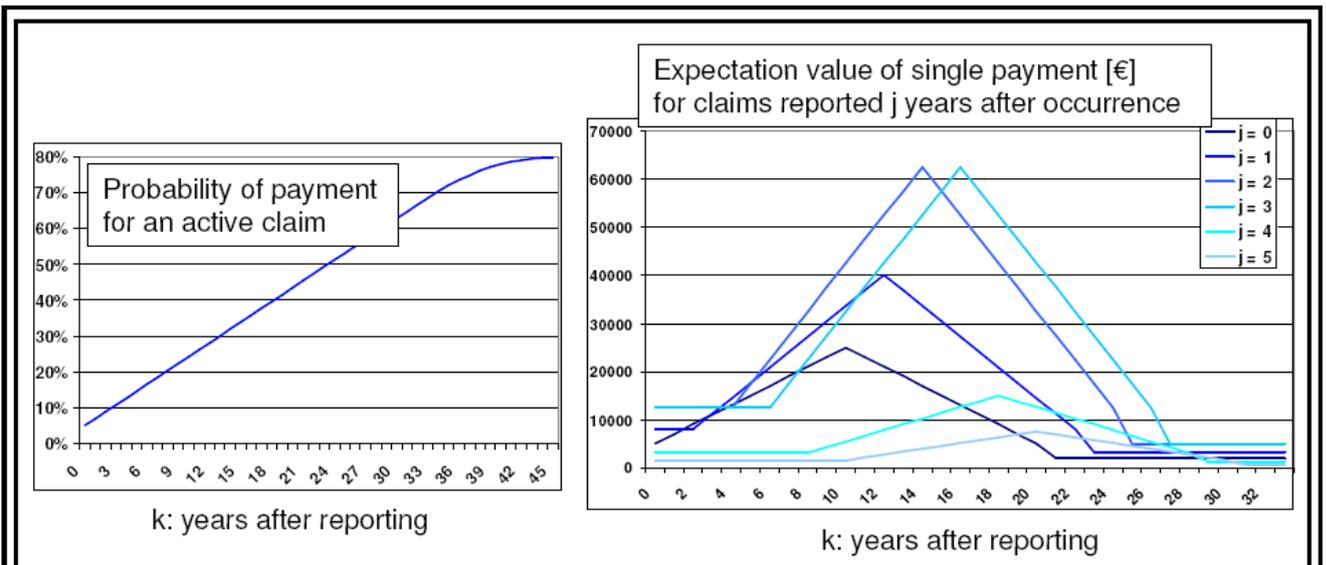

**Fig. 5:**
Model parameters: Probability of payment for an active claim (left) and Expectation value of single payment (right) for third party liability insurance / person injuries

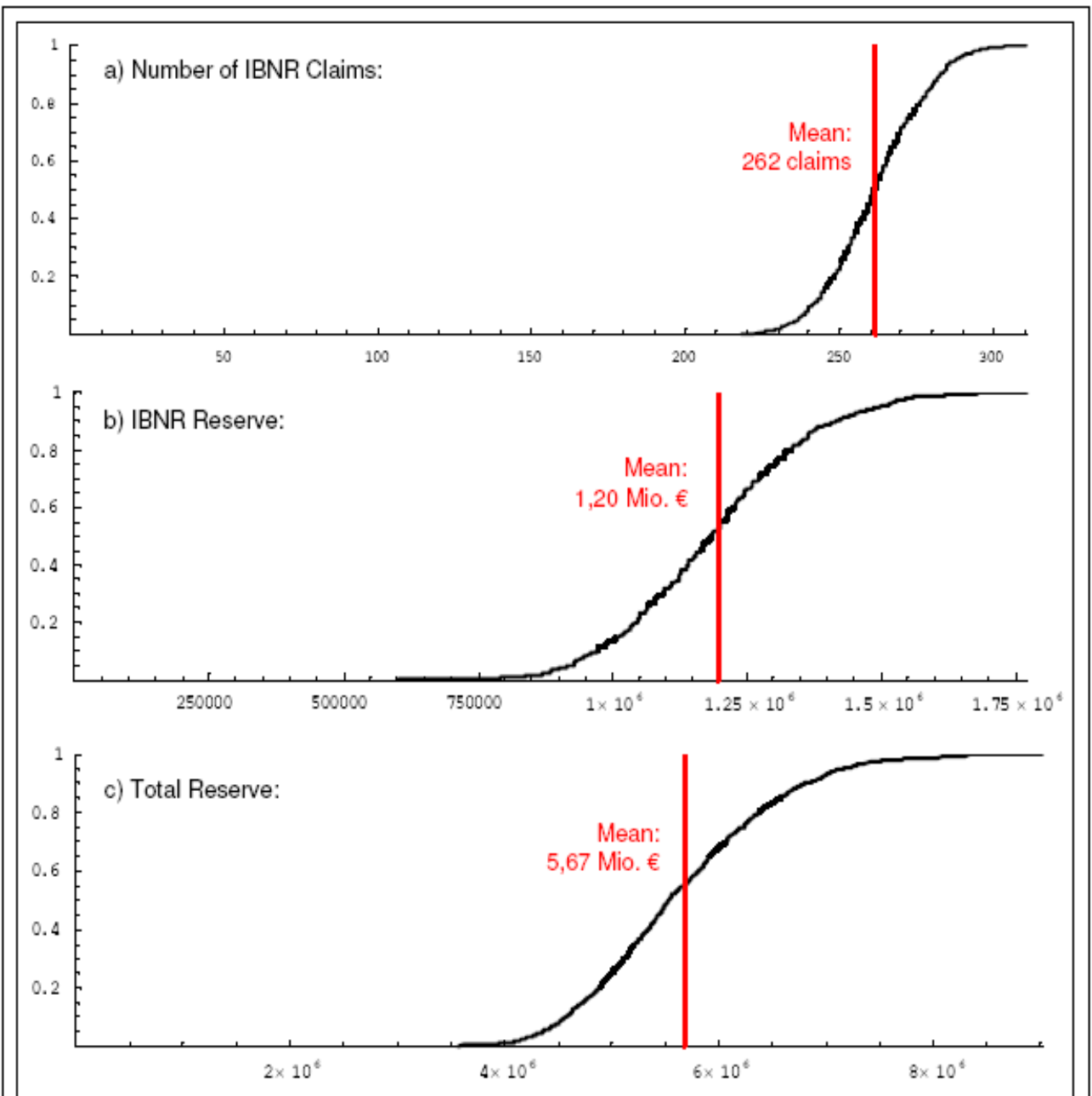

Fig. 6:
Result of MC Simulation: Empirical distribution functions of
a) Number of IBNR Claims
b) IBNR Reserve
c) Total Reserve
Each graph with the means according to the parameter values